\documentclass[twocolumn,conference]{IEEEtran}
\usepackage[T1]{fontenc}
\usepackage{prettyref}
\usepackage{amsmath}
\usepackage{amsthm}
\usepackage{amssymb}
\usepackage{graphicx}

\makeatletter
\theoremstyle{plain}
\newtheorem{thm}{\protect\theoremname}

\usepackage[caption=false,font=footnotesize]{subfig}

\makeatother

\providecommand{\theoremname}{Theorem}

\begin{document}

\title{On Energy Efficiency of the Nearest-Neighbor Cooperative Communication
in Heterogeneous Networks}

\author{\IEEEauthorblockN{Tao~Han\IEEEauthorrefmark{1}, Yu~Feng\IEEEauthorrefmark{1}, Jiang~Wang\IEEEauthorrefmark{2}\IEEEauthorrefmark{3},
Lijun~Wang\IEEEauthorrefmark{4}, Qiang~Li\IEEEauthorrefmark{1}
and Yujie Han\IEEEauthorrefmark{1}}\IEEEauthorblockA{\IEEEauthorrefmark{1}School of Electronic Information and Communications,
Huazhong University of Science and Technology, Wuhan, China}\IEEEauthorblockA{\IEEEauthorrefmark{2}Shanghai Research Center for Wireless Communication,
Shanghai, China}\IEEEauthorblockA{\IEEEauthorrefmark{3}Shanghai Institute of Microsystem and Information
Technology, Chinese Academy of Sciences, Shanghai, China}\IEEEauthorblockA{\IEEEauthorrefmark{4}Department of Information Science and Technology,
Wenhua College, Wuhan, China}\IEEEauthorblockA{Email: \IEEEauthorrefmark{1}\{hantao, m201371739, qli\_patrick, m201471912\}@hust.edu.cn,
\IEEEauthorrefmark{2}jiang.wang@wico.sh, \IEEEauthorrefmark{4}wanglj22@163.com}}
\maketitle
\begin{abstract}
In this paper, we consider a two-dimensional heterogeneous cellular
network scenario consisting of one base station (BS) and some mobile
stations (MSs) whose locations follow a Poisson point process (PPP).
The MSs are equipped with multiple radio access interfaces including
a cellular access interface and at least one short-range communication
interface. We propose a nearest-neighbor cooperation communication
(NNCC) scheme by exploiting the short-range communication between
a MS and its nearest neighbor to collaborate on their uplink transmissions.
In the proposed cooperation scheme, a MS and its nearest neighbor
first exchange data by the short-range communication. Upon successful
decoding of the data from each other, they proceed to send their own
data, as well as the data received from the other to the BS respectively
in orthogonal time slots. The energy efficiency analysis for the proposed
scheme is presented based on the characteristics of the PPP and the
Rayleigh fading channel. Numerical results show that the NNCC scheme
significantly improves the energy efficiency compared to the conventional
non-cooperative uplink transmissions.\footnote{The corresponding author is Lijun Wang. The authors would like to
acknowledge the support from the International Science and Technology
Cooperation Program of China (Grant No. 2014DFA11640, 2012DFG12250
and 0903), the National Natural Science Foundation of China (NSFC)
(Grant No. 61471180, 61271224 and 61301128), the NSFC Major International
Joint Research Project (Grant No. 61210002), the Ministry of Science
and Technology 863 program (Grant No. 2014AA01A707), the Hubei Provincial
Science and Technology Department (Grant No. 2013CFB188), the Fundamental
Research Funds for the Central Universities (Grant No. 2013ZZGH009
2013QN136, and 2014QN155), and Special Research Fund for the Doctoral
Program of Higher Education (Grant No. 20130142120044), and EU FP7-PEOPLE-IRSES
(Contract/Grant No. 247083, 318992 and 610524).}\end{abstract}

\begin{IEEEkeywords}
Cooperative communication; Poisson point process; heterogeneous cellular
network
\end{IEEEkeywords}

\section{Introduction}

Nowadays many of mobile stations (MSs), e.g, smart cellular phones,
tablets and PADs, are equipped with multiple radio access interfaces,
e.g, cellular radio access, wireless local area network (WLAN), Bluetooth
interfaces and etc.. As multi-mode MSs, they can constitute heterogeneous
cellular networks (HCNs) and make it possible to improve the performance
of cellular uplinks by serving as a relay to their neighboring MSs.
By some of short-range communication methods provided by the multi-mode
MSs, they can communicate with each other with significantly high
efficiency and quite low cost. Then the MSs can exploit the short-range
communication links among them along with the uplinks to the base
station (BS) to form the cooperative communication, which can improve
the performance of the HCNs with regard to rate, outage probability,
coverage, energy efficiency and etc.. This paper focuses on the improvement
of the energy efficiency of uplink cellular communications based on
cooperation between neighboring MSs in HCNs.

Cooperative diversity has already emerged as a new and effective technique
to combat fading and to decrease energy consumption in wireless networks.
The nearest neighbor relay scheme that relay is chosen to be the nearest-neighbor
to the user towards the BS (access-point) always has been applied
in \cite{SAK2010Relay12,ALT2012Balance14}. \cite{SAK2010Relay12}
proposes and analyzes the performance of two schemes: a distributed
nearest-neighbor relay assignment in which users can act as relays,
and an infrastructure-based relay assignment in which fixed relay
nodes are deployed in the network to help the users forward their
data. \cite{ALT2012Balance14} explores the balance between cooperation
through relay nodes and aggregated interference generation in large
decentralized wireless networks using decode-and-forward by the nearest
neighbor relay scheme. \cite{ESM2010EnergyEfficient13} proposes an
energy-efficient cooperative multicasting scheme by properly selecting
relay agents (RAs) based on their location, channel condition and
coverage. \cite{NB2013Contract15} studies the relay selection schemes
to reduce energy consumption, and the optimal number of cooperative
is also given. Besides, \cite{YU2010Channel18} is based on coded
cooperation, which combines cooperation and channel coding together.
To save bandwidth and improve the information transmission rate, network
coding \cite{Li2003tit4} is often used after the MSs receive each
other\textquoteright s information successfully. But based on some
criteria, \cite{ka2014Unicast17} finds more scenarios where network
coding has no gain on throughput or energy saving. Further more, many
existing works concentrate on the resource allocation in cooperative
networks. \cite{Adeane2006spawc6} presents both a centralized and
a distributed power allocation schemes to optimize the BER performance
of cooperative networks. To maximize the overall throughput, \cite{Zhang2013globecom7}
proposes an optimal power allocation. An adaptive coded cooperative
protocol based on incremental redundancy by a ACK/NACK feedback is
proposed in \cite{Alazem2008WiMob9}.

Many of the above works have presented valuable theories, methods
and technologies of cooperative communication. But there are still
some improvements left to perform. Zou \emph{et al}. in \cite{zou2013tcom}
investigate user terminals cooperating with each other in transmitting
their data packets to the BS, by exploiting the multiple network access
interfaces, which is called inter-network cooperation. Given a target
outage probability and data rate requirements, they analyze the energy
consumption of conventional schemes as compared to the proposed inter-network
cooperation. The results show that the inter-network cooperation can
significantly improve the energy efficiency of the uplink cellular
communications. However in practical view there are some limits of
this scheme work. In \cite{zou2013tcom}, it is required that the
cooperative MSs have the same distances to the BS and know the instantaneous
fading coefficients of both the short-range communication channel
and each cellular channel to form an orthogonal matrix used in cooperative.
In contrast, our paper intends to propose a more general and yet efficient
cooperative scheme for HCNs, in which the cooperative MSs do not need
to locate at the same distance away from the BS and are able to perform
the cooperative communication without knowing the channel status.

The main contributions of this paper are summarized as follows. At
first, we present a nearest-neighbor cooperative communication (NNCC)
scheme in a HCN consisting of different radio access networks, i.e.,
a short-range communication network and a cellular network. Then we
compare the proposed NNCC scheme to conventional schemes without user
cooperation under target outage probability and data rate requirements.
Secondly, we derive the energy efficiency of NNCC scheme in a Rayleigh
fading environment. Further more, given a target outage probability,
data rate requirements and distances between MSs and the BS, we derive
the cumulative distribution function (CDF) and the probability density
function (PDF) of energy consumption by considering the MSs that follow
a Poisson point process (PPP).

The remainder of this paper is organized as follows. Section \ref{sec:Cellular-Uplink-Transmission}
presents the network model and the NNCC scheme. In Section \ref{sec:Energy-Consumption-Analysis},
we present the desired power consumption analysis, then we derive
the CDF and the PDF of the system desired power consumption by considering
stochastic spatial distribution of cooperative MSs. Section \ref{sec:Numerical-Results}
gives the numerical results. Finally, Section \ref{sec:Conclusions}
concludes the paper.

\section{System Model\label{sec:Cellular-Uplink-Transmission}}

In this section, we first present a two-dimensional network model
of a HCN environment. Then, we propose a NNCC scheme by exploiting
the short-range network to assist cellular uplink transmissions.

\begin{figure}[tbh]
\begin{centering}
\includegraphics[width=0.75\columnwidth]{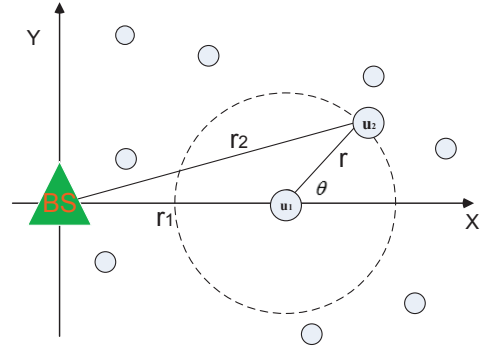}
\par\end{centering}

\raggedright{}\caption{\label{fig:System Model}System Model}
\end{figure}

\subsection{Network Model}

Consider a HCN consisting of a BS and some MSs whose locations follow
a homogeneous Poisson point process (PPP) with density $\rho$. The
MSs are assumed to equip with multiple radio access interfaces including
at least a short-range communication interface and a cellular access
interface. The MSs can communicate with their neighboring MSs by the
short-range communication. The packet size of data is assumed to be
the same across cellular communication link and all short-range communication
links between MSs. Without loss of generality, we consider the BS
is located at the origin of coordinate, and a specific MS U1 located
at coordinate $\left(r_{0},0\right)$ intends to communicate with
the BS. Our model is shown in Fig. \ref{fig:System Model}, U1 will
choose its nearest neighboring MS, denoted by U2, to cooperatively
communicate with the BS. According to the properties of the homogeneous
PPP, the distance $r$ between U1 and U2 satisfies the following PDF
\cite{stoyan1995stochastic} 
\begin{equation}
f_{r}(r)=2\pi\rho r\exp\left(-\pi\rho r^{2}\right),\label{eq:fr}
\end{equation}
then the coordinate of U2 can be obtained as $\left(r_{0}+r\cos\theta,r\sin\theta\right)$,
where $\theta$ follows a uniform distribution between $-\frac{\pi}{2}$
and $\frac{3\pi}{2}$. Denote the distance between U1 and the BS by
$r_{1}$. The distance $r_{\mathrm{2}}$ between the MS U2 and BS
can be obtained as

\begin{equation}
r_{\mathrm{2}}^{2}=r{}^{2}+r_{1}^{2}+2r_{1}r\text{\ensuremath{\cos}}\left(-\theta\right)=r{}^{2}+r_{1}^{2}+2r_{1}r\cos\theta.\label{eq:d2b}
\end{equation}

We consider a general channel model that incorporates the radio frequency,
path loss and fading effects in characterizing wireless transmissions,
i.e.,

\begin{equation}
\mathcal{P}_{{\rm \mathrm{R}}}=\mathcal{P}_{{\rm \mathrm{T}}}\left(\frac{\lambda}{4\pi d}\right)^{2}G_{{\rm \mathrm{T}}}G_{{\rm \mathrm{R}}}\left|h\right|^{2},\label{eq:pt-pr}
\end{equation}
 where $\mathcal{P}_{\mathrm{R}}$ is the received power, $\mathcal{P}_{\mathrm{T}}$
is the transmitted power, $\lambda$ is the carrier wavelength, $d$
is the transmission distance, $G_{\mathrm{T}}$ is the transmit antenna
gain, $G_{\mathrm{R}}$ is the receive antenna gain, and $h$ is the
channel fading coefficient. In this paper, we consider a Rayleigh
fading model to characterize the channel fading, i.e., $|h|^{2}$
is modeled as an exponential random variable.

\subsection{The NNCC Scheme}

\begin{figure}[tbh]
\begin{centering}
\includegraphics[width=0.7\columnwidth]{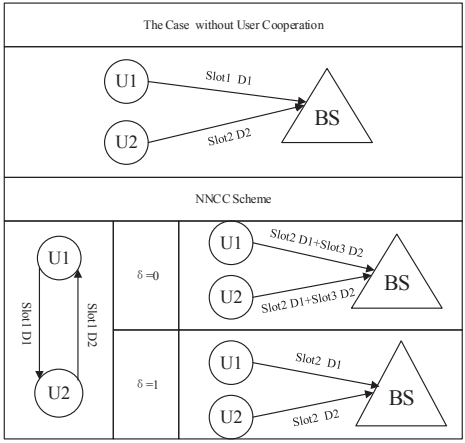}
\par\end{centering}

\raggedright{}\caption{\label{fig:NNCC}In NNCC scheme, transmissions happen in short range
network channel in time slot 1, and in cellular network channel in
time slot 2 (and time slot 3).}
\end{figure}

In the NNCC scheme, when MS U1 and its nearest neighbor MS U2 intend
to send data $D_{1}$ and $D_{2}$ to the BS, respectively, they cooperate
with each other according to the following steps:
\begin{enumerate}
\item U1 and U2 exchange their data over the short-range communication network
in time slot 1.
\item If both U1 and U2 succeed in decoding the data from each other, defined
as the case $\delta=0$, both of them will send their own data as
well as the data received from the other side to the BS in time slot
2 and time slot 3, respectively, i.e., they send both $D_{1}$ and
$D_{2}$ to the BS over two orthogonal cellular uplink channels. Otherwise,
defined as the case $\delta=1$, U1 and U2 will send only their own
data to the BS separately in time slot 2, just like a conventional
non-cooperation communication.
\end{enumerate}
Assuming that the short-range channels among MSs and the cellular
channels to the BS are orthotropic and there is no interference at
the BS among the MSs' signals, we only consider the channel noise
when analyzing the performance of the scheme. Considering that Ui
transmits $D_{{\rm i}}$ to Uj with the signal power $\mathcal{P}_{ij}^{\mathrm{NC}}$,
we can obtain the received signal-to-noise-ratio (SNR) between MSs
by NNCC scheme as

\begin{equation}
\gamma_{\mathrm{ij}}^{\mathrm{NC}}=\frac{\mathcal{P}_{\mathrm{ij}}^{\mathrm{NC}}}{N_{0}B_{\mathrm{s}}}\left(\frac{\lambda_{\mathrm{s}}}{4\pi r}\right)^{2}G_{\mathrm{U1}}G_{\mathrm{U2}}\left|h_{\mathrm{ij}}\right|^{2},\label{eq:r12nc}
\end{equation}
where $\mathrm{i}=1\,\mathrm{or}\,2$, $\mathrm{j}=2\,\mathrm{or}\,1$,
${\rm i}\neq{\rm j},$ $\lambda_{s}$ is the carrier wavelength of
the short-range communication, $G_{{\rm U}1}$ is the antenna gain
at U1, $G_{\mathrm{U2}}$ is the antenna gain at U2, and $h_{{\rm ij}}$
is the fading coefficient of the channel from Ui to Uj. The noise
is modeled as $N_{0}B_{s}$, where $N_{0}$ is the noise power spectral
density and $B_{s}$ is the channel bandwidth.

In step 2, MSs will transmit the data to the BS, and the received
SNR at the BS from MS U1 or U2 over the cellular channel can be obtained
as

\begin{equation}
\gamma_{\mathrm{ib}}=\frac{\mathcal{P}_{\mathrm{ib}}}{N_{0}B_{\mathrm{c}}}\left(\frac{\lambda_{\mathrm{c}}}{4\pi r_{i}}\right)^{2}G_{\mathrm{Ui}}G_{\mathrm{BS}}\left|h_{\mathrm{ib}}\right|^{2},\label{eq:r1bnc1}
\end{equation}
where $\mathrm{i}=1\,\mathrm{or}\,2$, $\lambda_{c}$ is the cellular
carrier wavelength, $B_{c}$ is the cellular spectrum bandwidth, $G_{\mathrm{BS}}$
is the receive antenna gain at BS, and $h_{\mathrm{ib}}$ is the fading
coefficient of channel from Ui to BS.

\section{Energy Efficiency Analysis of NNCC scheme\label{sec:Energy-Consumption-Analysis}}

In this section, we analyze the energy efficiency of the proposed
NNCC scheme compared to the conventional scheme without cooperation,
under the requirements of target outage probability $P_{\mathrm{out}}$
and data rate $R$.

\subsection{Energy Consumption in the NNCC scheme}
\begin{thm}[]
\label{thm:NNCC_energy_consumpution}Under the situation that the
BS succeeds in receiving the complete data from both MSs, the power
consumption for the NNCC scheme can be obtained as
\begin{equation}
\mathcal{P}^{\mathrm{NC}}=\mathcal{P}_{\mathrm{12}}^{\mathrm{NC}}+\mathcal{P}_{\mathrm{21}}^{\mathrm{NC}}+(1+(1-P_{{\rm out}})^{2})(\mathcal{P}_{\mathrm{1b}}^{\mathrm{NC}}+\mathcal{P}_{\mathrm{2b}}^{\mathrm{NC}}),\label{eq:pnc1}
\end{equation}
where \textup{$P_{{\rm out}}$ }is the target outage probability,\textup{
}and to meet the target outage probability, where $\mathcal{P}_{12}^{\mathrm{NC}}$,
$\mathcal{P}_{21}^{\mathrm{NC}}$, $\mathcal{P}_{1\mathrm{b}}^{\mathrm{NC}}$
and $\mathcal{P}_{2\mathrm{b}}^{\mathrm{NC}}$ are the desired transmission
power from U1 to U2, U2 to U1, U1 to the BS and U2 to the BS, respectively,
which are given by \eqref{eq:p12nc} and \eqref{eq:p1bnc1}.\end{thm}
\begin{IEEEproof}
Due to the limited error correction capability in practical communication
systems, both the short-range and cellular communications cannot achieve
the Shannon capacity. Therefore, let $\Delta_{\mathrm{s}}>1$ and
$\Delta_{\mathrm{c}}>1$ denote the performance gaps for the short-range
communication and the cellular communication from their respective
capacity limits, respectively. Using \eqref{eq:pt-pr} and considering
the performance gap $\Delta_{\mathrm{s}}$ away from Shannon capacity,
we obtain the maximum achievable rate from U1 to U2 of the short-range
communication of the NNCC scheme as

\begin{align}
C_{\mathrm{12}}^{\mathrm{NC}} & =B_{\mathrm{s}}\log_{2}(1+\frac{\gamma_{\mathrm{12}}^{\mathrm{NC}}}{\Delta_{\mathrm{s}}})\nonumber \\
 & =B_{\mathrm{s}}\log_{2}\left(1+\frac{\mathcal{P}_{\mathrm{12}}^{\mathrm{NC}}G_{\mathrm{U1}}G_{\mathrm{U2}}\left|h_{12}\right|^{2}}{\Delta_{\mathrm{s}}N_{0}B_{\mathrm{s}}}\left(\frac{\lambda_{\mathrm{s}}}{4\pi r}\right)^{2}\right).\label{eq:c12NC}
\end{align}

In a Rayleigh fading channel, all random variables $\left|h_{12}\right|^{2}$,
$\left|h_{21}\right|^{2}$, $\left|h_{\mathrm{1b}}\right|^{2}$ and
$\left|h_{\mathrm{2b}}\right|^{2}$ follow independent exponential
distributions with means $\sigma_{12}^{2}$, $\sigma_{21}^{2}$, $\sigma_{\mathrm{1b}}^{2}$
and $\sigma_{\mathrm{2b}}^{2}$, respectively. As we know, an outage
event occurs when the channel capacity falls below the required data
rate. Using \eqref{eq:r12nc} and considering the performance gap
$\Delta_{\mathrm{s}}$ away from Shannon capacity, we can obtain the
outage probability of the short-range transmission from U1 to U2 as

\begin{align}
P_{\mathrm{out12}}^{\mathrm{NC}} & =\Pr\left(C_{\mathrm{12}}^{\mathrm{NC}}<R\right)\nonumber \\
 & =1-\exp\left(-\frac{16\pi^{2}\triangle_{\mathrm{s}}N_{0}B_{\mathrm{s}}r^{2}\left(2^{\frac{R}{B_{\mathrm{s}}}}-1\right)}{\mathcal{P}_{\mathrm{12}}^{\mathrm{NC}}\sigma_{\mathrm{12}}^{2}G_{\mathrm{U1}}G_{\mathrm{U2}}\lambda_{\mathrm{s}}^{2}}\right).\label{eq:pout12nc}
\end{align}

Assuming $P_{\mathrm{out12}}^{\mathrm{NC}}=P_{\mathrm{out}}$, we
can obtain the desired power consumption of MSs for short-range communication
$\mathcal{P}_{\mathrm{ij}}^{\mathrm{NC}}$ from \eqref{eq:pout12nc}
as
\begin{equation}
\mathcal{P}_{\mathrm{ij}}^{\mathrm{NC}}=-\frac{16\pi^{2}\triangle_{\mathrm{s}}N_{0}B_{\mathrm{s}}\left(2^{\frac{R}{B_{\mathrm{s}}}}-1\right)}{\sigma_{\mathrm{ij}}^{2}G_{\mathrm{U1}}G_{\mathrm{U2}}\lambda_{\mathrm{s}}^{2}\ln\left(1-P_{\mathrm{out}}\right)}r^{2}.\label{eq:p12nc}
\end{equation}

Given $\sigma_{\mathrm{21}}^{2}=\sigma_{\mathrm{12}}^{2}$, we can
obtain $\mathcal{P}_{\mathrm{12}}^{\mathrm{NC}}=\mathcal{P}_{\mathrm{21}}^{\mathrm{NC}}=\zeta r^{2}$,
where $\zeta=-\frac{16\pi^{2}\triangle_{\mathrm{s}}N_{0}B_{\mathrm{s}}\left(2^{\frac{R}{B_{\mathrm{s}}}}-1\right)}{\sigma_{\mathrm{ij}}^{2}G_{\mathrm{U1}}G_{\mathrm{U2}}\lambda_{\mathrm{s}}^{2}\ln\left(1-P_{\mathrm{out}}\right)}$.

As discussed before, case $\delta=0$ implies that both U1 and U2
succeed in decoding each other\textquoteright s signals through short
range communications, and $\delta=1$ means that either U1 or U2 (or
both) fails to decode in the short-range transmissions. We can describe
$\delta=0$ and $\delta=1$ as follows.

$\delta=0$:
\begin{equation}
B_{\mathrm{s}}\log_{2}\left(1+\frac{\gamma_{12}^{\mathrm{NC}}}{\Delta_{\mathrm{s}}}\right)\geqslant R\,{\rm and}\,B_{\mathrm{s}}\log_{2}\left(1+\frac{\gamma_{21}^{\mathrm{NC}}}{\Delta_{\mathrm{s}}}\right)\geqslant R.\label{eq:xita0}
\end{equation}

$\delta=1:$
\begin{equation}
B_{\mathrm{s}}\log_{2}\left(1+\frac{\gamma_{12}^{\mathrm{NC}}}{\Delta_{\mathrm{s}}}\right)<R\,{\rm or}\,B_{\mathrm{s}}\log_{2}\left(1+\frac{\gamma_{21}^{\mathrm{NC}}}{\Delta_{\mathrm{s}}}\right)<R.\label{eq:xita1}
\end{equation}

Denote the target outage probability for short-range communication
between U1 and U2 by $P_{{\rm out}}$, given $P_{\mathrm{out}1\mathrm{2}}^{\mathrm{NC}}=P_{\mathrm{out21}}^{\mathrm{NC}}=P_{\mathrm{out}}$,
we have
\[
\Pr\left(\delta=0\right)=\left(1-P_{{\rm out}}\right)^{2},
\]
and
\[
\Pr\left(\delta=1\right)=1-\left(1-P_{{\rm out}}\right)^{2}.
\]

Moreover, denote the target outage probability for cellular communication
from U1, U2 to the BS by $P_{\mathrm{out}}^{{\rm NC}}$, and given
$P_{\mathrm{out}1\mathrm{b}}^{\mathrm{NC}}=P_{\mathrm{out2b}}^{\mathrm{NC}}=P_{\mathrm{out}}^{{\rm NC}}$,
we have $\Pr\left(C_{\mathrm{1b}}<R\right)=\Pr\left(C_{\mathrm{2b}}<R\right)$,
then we obtain the outage probability of the NNCC scheme by from \eqref{eq:xita0}
and \eqref{eq:xita1} as

$P_{\mathrm{out}}=\left(1-P_{{\rm out}}\right)^{2}*\left(P_{\mathrm{out}}^{{\rm NC}}\right)^{2}+\left(1-\left(1-P_{{\rm out}}\right)^{2}\right)*\left(1-\left(1-P_{\mathrm{out}}^{{\rm NC}}\right)^{2}\right)$.

Then, we obtain
\begin{equation}
P_{\mathrm{out}}^{{\rm NC}}=\frac{\sqrt{\left(1-\epsilon\right)^{2}+P_{\mathrm{out}}\left(2\epsilon-1\right)}-\left(1-\epsilon\right)}{2\epsilon-1}\text{,}\label{eq:p11}
\end{equation}
where $\epsilon=\left(1-P_{{\rm out}}\right)^{2}$. 

We can obtain the power consumption of Ui for cellular communication
from \eqref{eq:p11} as
\begin{equation}
\mathcal{P}_{\mathrm{ib}}^{\mathrm{NC}}=-\frac{16\pi^{2}\triangle_{\mathrm{c}}N_{0}B_{\mathrm{c}}\left(2^{\frac{R}{B_{\mathrm{c}}}}-1\right)}{\sigma_{\mathrm{ib}}^{2}G_{\mathrm{Ui}}G_{\mathrm{BS}}\lambda_{\mathrm{c}}^{2}\ln\left(1-P_{\mathrm{out}}^{{\rm NC}}\right)}r_{i}^{2}.\label{eq:p1bnc1}
\end{equation}

Given $\sigma_{\mathrm{1b}}^{2}=\sigma_{\mathrm{2b}}^{2}$, we can
obtain $\mathcal{P}_{\mathrm{1b}}^{\mathrm{NC}}=\eta r_{1}^{2}$ and
$\mathcal{P}_{\mathrm{2b}}^{\mathrm{NC}}=\eta r_{2}^{2}$, where $\eta=-\frac{16\pi^{2}\triangle_{\mathrm{c}}N_{0}B_{\mathrm{c}}\left(2^{\frac{R}{B_{\mathrm{c}}}}-1\right)}{\sigma_{\mathrm{ib}}^{2}G_{\mathrm{U2}}G_{\mathrm{BS}}\lambda_{\mathrm{c}}^{2}\ln\left(1-P_{\mathrm{out}}^{{\rm NC}}\right)}$.

Notice that in case of $\delta=0$, cooperation communication is employed
and there are energy consumption at both time slot 2 and time slot
3, resulting in that a total power consumption of $2(\mathcal{P}_{\mathrm{1b}}^{\mathrm{NC}}+\mathcal{P}_{\mathrm{2b}}^{\mathrm{NC}})$
is consumed by U1 and U2 in transmitting to BS. In case of $\delta=1$,
U1 and U2 consume a total power consumption of $(\mathcal{P}_{\mathrm{1b}}^{\mathrm{NC}}+\mathcal{P}_{\mathrm{2b}}^{\mathrm{NC}})$
for transmitting to BS. Therefore, considering both the short-range
communication and cellular transmissions, the total power consumption
by the NNCC scheme is given by
\begin{align*}
 & \mathcal{P}^{\mathrm{NC}}\\
 & =\mathcal{P}_{\mathrm{12}}^{\mathrm{NC}}+\mathcal{P}_{\mathrm{21}}^{\mathrm{NC}}+\left(2Pr\left(\delta=0\right)+Pr\left(\delta=1\right)\right)(\mathcal{P}_{\mathrm{1b}}^{\mathrm{NC}}+\mathcal{P}_{\mathrm{2b}}^{\mathrm{NC}})\\
 & =\mathcal{P}_{\mathrm{12}}^{\mathrm{NC}}+\mathcal{P}_{\mathrm{21}}^{\mathrm{NC}}+(1+(1-P_{{\rm out}})^{2})(\mathcal{P}_{\mathrm{1b}}^{\mathrm{NC}}+\mathcal{P}_{\mathrm{2b}}^{\mathrm{NC}}).
\end{align*}

\end{IEEEproof}

In order to compare our method and the traditional method, we give
the same target interrupt probability definition about two schemes,
and then we will derive the power consumption for the conventional
scheme without user cooperation, in which both U1 and U2 succeed in
transmitting their data to the BS separately.  Similarly to the power
consumption analysis of the NNCC scheme, assuming that $P_{\mathrm{out1b}}^{\mathrm{C}}=P_{\mathrm{out2b}}^{\mathrm{C}}=P_{{\rm out}}^{{\rm C}}$,
the total power consumption of U1 and U2 under conventional non-cooperative
communication can be obtained by 
\begin{equation}
\mathcal{P}^{\mathrm{C}}=\sum_{i=1}^{2}\mathcal{P}_{\mathrm{ib}}^{\mathrm{C}},\label{eq:pt}
\end{equation}
 where 
\begin{equation}
\mathcal{P}_{\mathrm{ib}}^{\mathrm{C}}=-\frac{16\pi^{2}\triangle_{\mathrm{c}}N_{0}B_{\mathrm{c}}r_{\mathrm{i}}^{2}\left(2^{\frac{R}{B_{\mathrm{c}}}}-1\right)}{\sigma_{\mathrm{ib}}^{2}G_{\mathrm{Ui}}G_{\mathrm{BS}}\lambda_{\mathrm{c}}^{2}\ln\left(1-P_{{\rm out2}}^{{\rm C}}\right)},\label{eq:p1bt}
\end{equation}
 and 

\begin{equation}
P_{{\rm out}}^{{\rm C}}=1-\sqrt{1-P_{\mathrm{out}}}.\label{eq:pOUT12C}
\end{equation}

As \eqref{eq:p11} and \eqref{eq:pOUT12C} show, the NNCC scheme can
save more energy than the conventional scheme because it can work
under larger target outage probability. \prettyref{thm:NNCC_energy_consumpution}
and \prettyref{eq:pt} give the relations between the desired transmission
powers and the target outage probabilities as well as other impact
factors, e.g., path loss, fading, and thermal noise, under the NNCC
scheme and the conventional non-cooperative scheme, respectively.
Based on them, some further performance analysis, such as the energy
efficiency analysis in Section \prettyref{sub:Energy-Efficiency-Analysis},
can be presented.

\subsection{Energy Efficiency Analysis based on PPP\label{sub:Energy-Efficiency-Analysis} }

In order to get more performance analysis about the NNCC scheme,
we put the NNCC scheme on a more general environment, we will consider
MSs satisfy Poisson point process, which meet actual situation. The
result is more practical and performance analysis is more accurate.
Moreover, considering the stochastic spatial distribution of the MSs,
we can derive the CDF and PDF of the desired transmission power which
meet the target outage probability and rate requirement.
\begin{thm}[]
\label{thm:pdf_cdf_ee} When the spatial distribution of the MSs
follows a homogeneous PPP with density $\rho$, the PDF of \textup{$\mathrm{\mathcal{P}}^{\mathrm{NC}}$}
is \textup{
\[
f_{\mathrm{\mathrm{\mathcal{P}}^{\mathrm{NC}}}}\left(\mathrm{\mathrm{\mathcal{P}}^{\mathrm{NC}}}\right)=\begin{cases}
\intop_{\frac{\pi}{2}}^{\frac{3\pi}{2}}\left(\frac{\rho R_{2}\mathrm{e}^{\left(-\pi\rho\mathrm{R_{2}^{2}}\right)}+\rho R_{1}\mathrm{e}^{\left(-\pi\rho\mathrm{R_{1}^{2}}\right)}}{2\varDelta_{R}}\right)d\theta, & Q_{1}\\
\intop_{-\frac{\pi}{2}}^{\frac{3\pi}{2}}\frac{\rho R_{2}\mathrm{e}^{\left(-\pi\rho\mathrm{R_{2}^{2}}\right)}}{2\varDelta_{R}}d\theta, & Q_{2},
\end{cases}
\]
}and the CDF of \textup{$\mathrm{\mathcal{P}}^{\mathrm{NC}}$} is
\textup{
\[
F_{\mathrm{\mathrm{\mathcal{P}}^{\mathrm{NC}}}}\left(\mathrm{\mathrm{\mathcal{P}}^{\mathrm{NC}}}\right)=\begin{cases}
\intop_{\frac{\pi}{2}}^{\frac{3\pi}{2}}\frac{\mathrm{e}^{\left(-\pi\rho\mathrm{R_{1}^{2}}\right)}-\mathrm{e}^{\left(-\pi\rho\mathrm{R_{2}^{2}}\right)}}{2\pi}d\theta, & Q_{1}\\
\intop_{-\frac{\pi}{2}}^{\frac{3\pi}{2}}\frac{1-\mathrm{e}^{\left(-\pi\rho\mathrm{R_{2}^{2}}\right)}}{2\pi}d\theta+F_{\mathrm{\mathcal{\mathrm{\mathcal{P}}^{\mathrm{NC}}}}}\left(\mathrm{2\varepsilon\eta r_{1}^{2}}\right), & Q_{2}\text{,}
\end{cases}
\]
}where 
\begin{equation}
\varepsilon=1+(1-P_{{\rm out}})^{2}\text{,}\label{eq:I}
\end{equation}
\textup{
\begin{equation}
\varDelta_{R}=\sqrt{(\varepsilon\eta r_{1}\cos\theta)^{2}-(2\zeta+\varepsilon\eta)(2\varepsilon\eta r_{1}^{2}-\mathcal{P}^{\mathrm{NC}})},\label{eq:panbieshi}
\end{equation}
}
\begin{equation}
R_{2}=\frac{-\varepsilon\eta r_{1}\cos\theta+\varDelta_{R}}{2\zeta+\varepsilon\eta},\label{eq:r2}
\end{equation}
and 
\begin{equation}
R_{1}=\frac{-\varepsilon\eta r_{1}\cos\theta-\varDelta_{R}}{2\zeta+\varepsilon\eta}.\label{eq:r1}
\end{equation}
 $Q_{1}$ stands for \textup{$2\varepsilon\eta r_{1}^{2}-\frac{\left(\varepsilon\eta r_{1}\cos\theta\right)^{2}}{2\zeta+\varepsilon\eta}<\mathrm{\mathcal{P}}^{\mathrm{NC}}\leqslant2\varepsilon\eta r_{1}^{2}$}
and $Q_{2}$ stands for \textup{$\mathrm{\mathcal{P}}^{\mathrm{NC}}>2\varepsilon\eta r_{1}^{2}.$} \end{thm}
\begin{IEEEproof}
Due to independence of random variables $r$ and $\theta$, considering
$\theta$ follows the uniform distribution between $-\frac{\pi}{2}$
and $\frac{3\pi}{2}$, $r$ satisfies \eqref{eq:fr}, the PDF of $r$
and $\theta$ can be obtained as 
\begin{equation}
f(r,\theta)=\rho r\exp\left(-\pi\rho r^{2}\right).\label{eq:frjiao-1-1-1}
\end{equation}
Substituting \eqref{eq:pnc1} into \eqref{eq:d2b}, we obtain
\begin{equation}
\mathrm{\mathcal{P}}^{\mathrm{NC}}=\left(2\zeta+\varepsilon\eta\right)r^{2}+2\varepsilon\eta r_{1}\cos\theta r+2\varepsilon\eta r_{1}^{2}.\label{eq:grjiao}
\end{equation}
 So the the CDF of $\mathrm{\mathcal{P}}^{\mathrm{NC}}$ is

\begin{align}
 & F_{\mathcal{P}^{\mathrm{NC}}}\left(p^{{\rm NC}}\right)=\Pr\left(P^{{\rm NC}}\leqslant p^{{\rm NC}}\right)\nonumber \\
 & =\Pr\left(\left(2\zeta+\varepsilon\eta\right)r^{2}+2\varepsilon\eta r_{1}\cos\theta r+2\varepsilon\eta r_{1}^{2}\leqslant p^{{\rm NC}}\right).\label{eq:cdf_pnc}
\end{align}
 Based on \eqref{eq:frjiao-1-1-1} and \eqref{eq:cdf_pnc}, we can
acquire the CDF and PDF of desired power for NNCC scheme as \prettyref{thm:pdf_cdf_ee}
expresses.
\end{IEEEproof}
From \prettyref{thm:pdf_cdf_ee}, we will know the PDF and CDF of
$\mathrm{\mathcal{P}}^{\mathrm{NC}}$ and can easily obtain its expectation.
It can help us to find the energy efficiency. Based on \prettyref{thm:pdf_cdf_ee},
it is easy to derive the relations between desired transmission power
of the NNCC scheme and the target outage probability as well as other
impact factors. Considering a successful transmission delivering both
data of U1 and U2 by the total power consumption $\mathrm{\mathcal{P}}^{\mathrm{NC}}$,
the energy efficiency of the NNCC scheme can be derived by 
\begin{equation}
EE^{\mathrm{NC}}=\frac{2R}{E\left(\mathrm{\mathrm{\mathcal{P}}^{\mathrm{NC}}}\right)}.\label{eq:ENERGY EFFICIENCY}
\end{equation}

In \prettyref{sec:Numerical-Results}, some numerical results about
the desired transmission power and the energy efficiency of the NNCC
scheme are given.

\section{Numerical Results\label{sec:Numerical-Results}}

In this section, we present the analytical results of the proposed
NNCC scheme and compare it to the existed schemes. We compare different
schemes under the same definition of target outage probability. In
our simulation, the frequency and bandwidth of the short-range communication
are given as $f_{{\rm s}}=2.4{\rm \,GHz}$ and ${\rm B_{s}}=2{\rm \,MHz}$,
respectively. In the cellular communication they are given as $f_{{\rm c}}=2100{\rm \,MHz}$
and ${\rm B}_{{\rm c}}=5\,{\rm MHz}$, respectively. The antenna gains
of U1 and U2 are set as $G_{{\rm U1}}=G_{{\rm U2}}=0\,{\rm dB}$ and
the BS\textquoteright s antenna gain is set as $G_{{\rm BS}}=5\,{\rm dB}$.
The performance gaps $\Delta_{{\rm s}}$ and $\Delta_{{\rm c}}$ of
short-range and cellular communication are given by $\Delta_{{\rm s}}=4\,{\rm dB}$
and $\Delta_{{\rm c}}=2\,{\rm dB}$.

\begin{figure}[htbp]
\begin{centering}
\includegraphics[width=0.94\columnwidth]{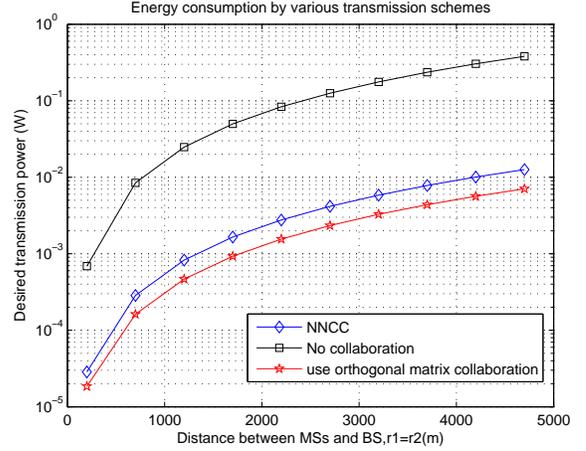}
\par\end{centering}

\caption{\label{fig:power changes with distance}Energy consumption by various
transmission schemes with target outage probability $P_{\mathrm{out}}=10^{-3}$,
effective rate $R=100000\,\mathrm{bits/s}$, and inter-user distance
$r=20\,\mathrm{m}$. }
\end{figure}

\begin{figure}[htbp]
\centering{}\includegraphics[width=0.94\columnwidth]{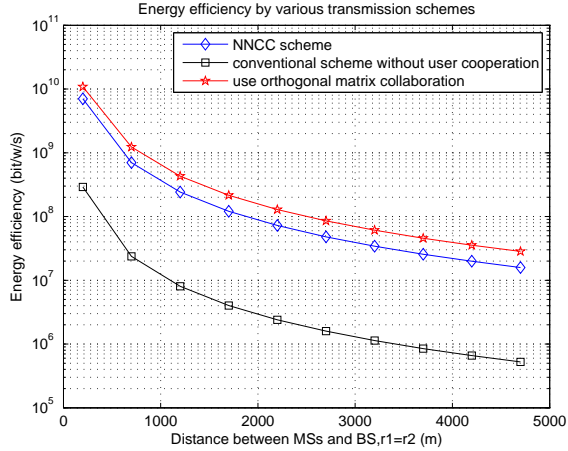}\caption{\label{fig:Energy efficiency two way}Energy efficiency by various
transmission schemes with target outage probability $P_{\mathrm{out}}=10^{-3}$,
effective rate $R=100000\mathrm{\,bits/s}$, and inter-user distance
$r=20\mathrm{\,m}$. }
\end{figure}

Energy consumption is influenced by some parameters, such as distances
between the MSs and the BS, distance between MSs, the density of MS
and the target outage probability. In Fig. \ref{fig:power changes with distance},
we discuss the impact of distances between the MSs and the BS on energy
consumption. We compare the desired transmission power of the NNCC
scheme to the conventional non-cooperative scheme and the inter-network
cooperative communication scheme by orthogonal matrix proposed in
\cite{zou2013tcom}. The result shows that the power consumption of
the NNCC scheme is significantly lower than the non-operative scheme
and slightly higher than the inter-network cooperative scheme. It
demonstrates the effectiveness of the NNCC scheme, while considering
the simplicity and feasibility of the NNCC scheme, which does not
need the knowledge of the instantaneous status of the channel and
can be easily applied in the scenario of the MSs with stochastic spatial
distribution. Similarly to Fig. \ref{fig:power changes with distance},
we present the comparison of the energy efficiency among the schemes
in Fig. \ref{fig:Energy efficiency two way}.

\begin{figure}[htbp]
\centering{}\includegraphics[width=0.94\columnwidth]{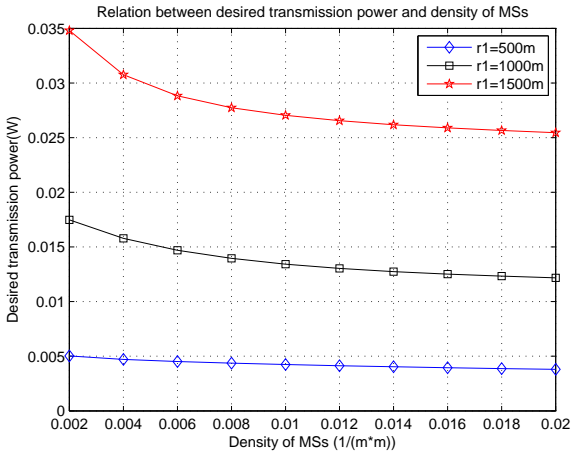}
\caption{\label{fig:power and density }Relation between the desired the transmission
power and the density of MSs, with target outage probability $P_{\mathrm{out}}=10^{-3}$,
effective rate $R=10000000\mathrm{\,bits/s}$, and U1-BS distance
$r_{1}=2000\mathrm{\,m}$. }
\end{figure}

\begin{figure}[htbp]
\centering{}\includegraphics[width=0.94\columnwidth]{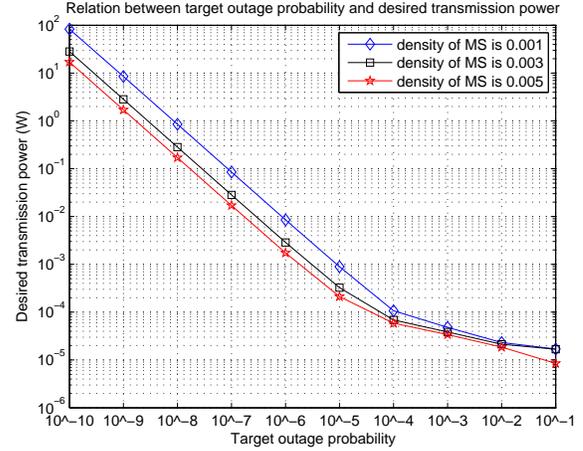}\caption{\label{fig:power and outage probability}Desired transmission power
with regard to the target outage probability with effective rate $R=1000000\,\mathrm{bits/s}$,
and U1-BS distance $r_{1}=150\mathrm{\,m}$.}
\end{figure}

In Fig. \ref{fig:power and density }, we present the relations between
the desired transmission power consumption and the density of the
MSs under different distances between the MS U1 and the BS. It is
shown that the power consumption decreases as the density of MSs increases,
because the distance between the cooperative MSs tends to be smaller
when the density of the MSs increases, and thus less power consumption
is required in step 1. Fig. \ref{fig:power and outage probability}
demonstrates that the energy consumptions of the NNCC scheme decreases
as the target outage probability increases with various densities
of the MSs.

\section{Conclusions\label{sec:Conclusions}}

In this paper, we propose a cooperative communication scheme, namely
the NNCC scheme, in which a MS in HCN and its nearest neighbor MS
exploit a short-range communication network to assist the cellular
transmissions. The energy efficiency of the NNCC scheme is analyzed
and derived in closed-form expressions, which provide insights into
the relations between energy efficiency and many important factors,
e.g., MS density, the distance between the MSs and the BS, the target
outage probability and etc.. Numerical results show that the NNCC
scheme is simple, yet efficient compared to the existing schemes.
Although this article studies collaboration between two MSs only,
it's obvious that there are vast energy saving comparing with the
traditional scheme, no matter from the aspects of the distance between
BS and MSs or expectation.   In this paper, we have only considered
the scenario of a single BS and allow only two MSs cooperate with
each other. In the future, we will extend the work to the multi-cell
scenario with multi-MS cooperative communications.

\appendices{}

\bibliographystyle{IEEEtran}
\bibliography{manu_icc2015-01}

\end{document}